\documentclass[aps,pra,notitlepage,%
longbibliography,twocolumn]{revtex4-2}
\usepackage{natbib}
\usepackage{graphicx}
\usepackage{amssymb}
\usepackage{amsmath}
\usepackage{ifthen}
\usepackage{braket}
\usepackage{bm}
\usepackage{bbm}
\usepackage{comment}
\usepackage{siunitx}
\usepackage{float}
\usepackage{dcolumn}
\newcolumntype{d}[1]{D{.}{.}{#1}}
\usepackage{subfigure}

\newcommand{\ii}{\mathrm{i}}

\newcommand{\hate}{\hat{\mathrm{e}}}

\newcommand{\calT}{\mathcal{T}}

\newcommand{\qq}{\ell}
\newcommand{\xy}{{\parallel}}

\newcommand{\plus}{{\mbox{\scriptsize $(+)$}}}
\newcommand{\minus}{{\mbox{\scriptsize $(-)$}}}

\begin{document}

\bibliographystyle{nsfbib}

\title{Radiative Energy and Mass Shifts of Quantum Cyclotron States}

\author{U. D. Jentschura}
\affiliation{Department of Physics and LAMOR, Missouri University of Science and
Technology, Rolla, Missouri 65409, USA}

\begin{abstract}
We discuss relativistic and radiative corrections 
to the energies of quantum cyclotron states.
In particular, it is shown analytically that 
the leading logarithmic radiative (self-energy) correction
to the bound-state energy levels of quantum cyclotron
states is state-independent, and must be interpreted
as a magnetic-field-dependent correction to the 
electron's mass in a Penning trap. 
\end{abstract}

\maketitle

%
%
\section{Introduction}
\label{sec1}

Quantum cyclotron states~\cite{BrGa1982,Br1985aop,BrGa1986} have a somewhat
unique position in the analysis of bound states
as the binding potential is externally induced,
via the electric and magnetic fields of a Penning trap. 
The quantum-mechanical problem can be 
separated, on the level of the nonrelativistic 
theory, into spin-flip, cyclotron, 
axial, and magnetron, motions~\cite{BrGa1982,Br1985aop,BrGa1986}.
Measurements of the anomalous magnetic moment
of the electron~\cite{BrGa1982,BrGaHeTa1985,GaEtAl2006everything,%
HaFoGa2008,FaGa2021prl,FaGa2021pra} 
proceed via a {\em spectroscopic approach} 
applied to the quantum bound states inside 
a Penning trap, hence the name {\em quantum cyclotron}.
The intricacies of the problem raise 
awareness of the relativistic and 
radiative corrections to quantum cyclotron 
states~\cite{WiMoJe2022,Je2023mag1,JeMo2023mag2}.

The problem is not without peculiarities:
Namely, the separability of the 
nonrelativistic quantum problem into the 
spin-flip, cyclotron, axial and 
magnetron motions is only realized 
after one establishes highly nontrivial 
commutation relations among the 
corresponding quantum-mechanical 
operators~\cite{BrGa1982,Br1985aop,BrGa1986}.
The calculation of relativistic corrections
necessitates the application of 
higher-order Foldy--Wouthuysen
transformations~\cite{FoWu1950} in 
a nonstandard physical system, where
the identification
of suitable coupling parameters (generalized
fine-structure constants pertaining 
to the cyclotron and axial motions, see
Ref.~\cite{WiMoJe2022}) helps in
obtaining consistent results.

For the treatment of the quantum electrodynamic
radiative corrections (notably, the self-energy),
it is necessary to employ fully relativistic 
Landau levels in the symmetric gauge~\cite{Je2023mag1},
which allow one to establish a
one-to-one correspondence between the 
nonrelativistic and relativistic bound states
of given cyclotron and magnetron quantum 
numbers~\cite{Je2023mag1}.

This brief paper is organized as follows.
In Sec.~\ref{sec2}, we present a brief review of 
the quantum numbers of the quantum cyclotron
problem, and point out a few idiosyncrasies
and peculiarities in the theoretical analysis.
In Sec.~\ref{sec3}, we present results for 
the leading relativistic corrections
to the bound-state quantum-cyclotron energies.
The independence of the spin-independent 
part of the leading radiative
correction (self-energy) of the quantum numbers 
is shown analytically in Sec.~\ref{sec4}.
The interpretation of the findings is left for 
the conclusions presented in Sec.~\ref{sec5}.
Units with $\hbar = c = \epsilon_0 = 1$ are
used throughout.

\begin{figure}[t!]
\begin{center}
\begin{minipage}{0.99\linewidth}
\begin{center}
\includegraphics[width=0.99\linewidth]{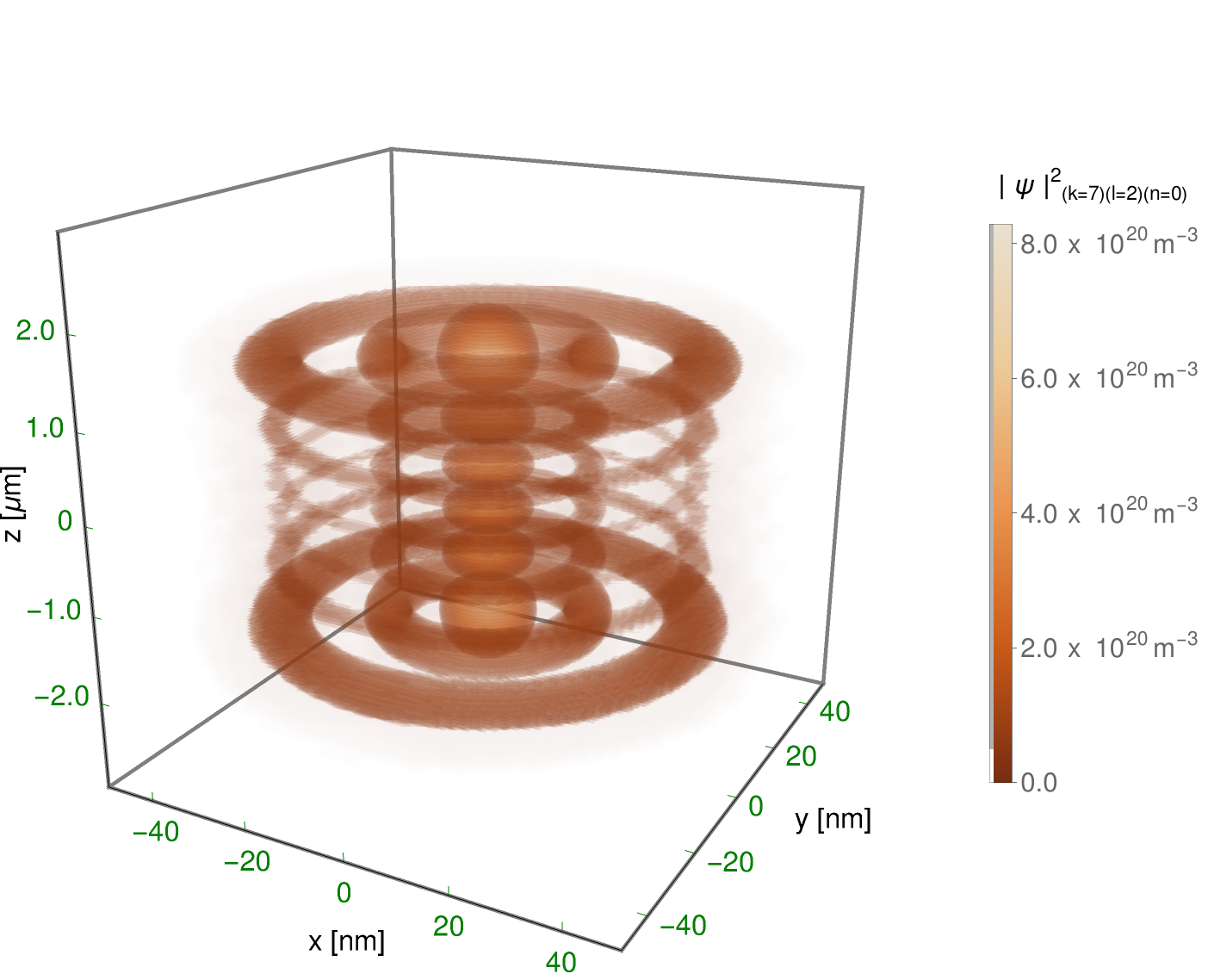}
\caption{\label{fig1}
We use parameters from Ref.~\cite{BrGa1986},
namely,
$\omega_c = 2 \pi \times 164.4 \, {\rm GHz}$,
$\omega_z = 2 \pi \times 64.42 \, {\rm MHz}$,
and $\omega_\minus = 2 \pi \times 12.62 \, {\rm kHz}$.
The probability density $| \psi |^2 = | \psi_{k \ell n s}(\vec r) |^2$
is plotted for the quantum cyclotron state with
quantum numbers $k=5$, $n=2$, and $\ell = 2$
[see Eq.~\eqref{psidef}].
The quantum numbers describe the fifth axial excited state ($k=7$),
the second excited cyclotron state ($n = 0$),
and the second excited magnetron state ($\ell = 2$).
Quantum mechanically, the second excited cyclotron state
has two nodes in the radial direction in the $xy$ plane.}
\end{center}
\end{minipage}
\end{center}
\end{figure}

%
%
\section{Quantum Numbers}
\label{sec2}

The quantum numbers characterizing 
a quantum cyclotron state are~\cite{BrGa1982,Br1985aop,BrGa1986} 
\begin{subequations}
\begin{align}
k =& \; 0,1,2,\dots \,, &
\qquad & \mbox{(axial)} \,,
\\[0.1133ex]
\qq =& \; 0,1,2,\dots \,, &
\qquad & \mbox{(magnetron)} \,,
\\[0.1133ex]
n =& \; 0,1,2,\dots &
\qquad & \mbox{(cyclotron)} \,,
\\[0.1133ex]
\label{s}
s =& \; \pm 1 \,,  &
\qquad & \mbox{(spin)} \,.
\end{align}
\end{subequations}
Here, the {\em axial} direction is along the $z$
axis, {\em i.e.}, in the direction of the magnetic 
field of the trap. The magnetic field
confines the motion of the bound electron 
in the $xy$ plane, to {\em cyclotron} orbits.
An additional electric quadrupole 
potential is proportional to a positive 
constant multiplying $z^2$ (in the $z$
direction), while it is multiplied
by a negative constant in the $xy$ plane
and leads to a repulsive potential 
proportional to $\rho^2 = x^2 + y^2$ in the 
$xy$ plane.

The quantum number $k$, $\ell$, $n$ and $s$ 
correspond to following aspects of the electron 
motion in the trap:
The axial quantum number $k$ describes the 
excitation of the harmonic-oscillator
states in the harmonic potential 
dependent on $z$.
The magnetic quantum number $\ell$ describes
a slow, circular drift motion around the 
center of the trap, which occurs due to 
a superposition of the Lorentz force from the 
trap's main magnetic field and the 
repulsive quadrupole potential (in the $xy$ plane).
The quantum number $n$ describes the (principal)
motion of the electron in the quantum cyclotron,
due to the binding magnetic field.
The quantum number $s$ describes the 
spin-flip of the electron in the binding field.
There is a certain idiosyncrasy 
in the notation: namely, the sequence
of quantum numbers follows alphabetical order:
$k$, then $\ell$, then $n$, then $s$. However, if we were to denote the
quantum numbers according to their ``importance'',
{\em i.e.}, with the quantum number having the 
largest effect on the spectrum used first, 
then in descending order, then the 
sequence would rather have to be 
$s \to n \to k \to \ell$. This is 
because the spin-flip frequency is larger than the 
cyclotron frequency due to the anomalous magnetic 
moment, and because the axial frequency, for typical
configurations~\cite{BrGa1986}, is 
a lot smaller than the cyclotron frequency,
while the magnetron frequency is again
a lot smaller than the axial one.

The nonrelativistic Hamiltonian can be written as the 
sum of three terms ($m$ is the electron mass), 
\begin{subequations}
\label{H0}
\begin{align}
H_0 =& \; H_\xy + H_\sigma + H_z \,,
\\[0.1133ex]
\label{H0rep2b}
H_\xy =& \; \frac{\vec p_\xy^{\,2}}{2m}
- \frac{e}{2m} \vec L \cdot \vec B
+ \frac{m \omega_c^2}{8} \,\rho^2 + V_\xy \,,
\\[0.1133ex]
H_\sigma =& \; - \frac{e}{2m} \,
(1 + \kappa) \, \vec\sigma\cdot \vec B \,,
\\[0.1133ex]
H_z =& \; \frac{p_z^2}{2 m} + V_z \,.
\end{align}
\end{subequations}
The cyclotron frequency is 
$\omega_c = |e| B/m$, where $e$ is the 
electron charge and $|e|$ its modulus.
Furthermore, $V_z = \frac12 m \omega_z^2 z^2$
is the axial potential,
where $\omega_z$ is the axial frequency,
while the repulsive 
quadrupole potential in the $xy$ plane 
is $V_\xy = - \frac14 m \omega_z^2 \rho^2 $,
where $\rho = \sqrt{x^2 + y^2}$.
The trap magnetic field is $\vec B = B \, \hate_z$.
Morevoer, $\kappa = \alpha/(2 \pi)$ is the 
one-loop Schwinger term describing the anomalous 
magnetic moment~\cite{Sc1948}.
The momentum operators are 
separated according to $\vec p_\parallel = p_x \hate_x + p_y \hate_y$,
while $p_z = - \ii \partial/\partial z$.

In the conventions of Sec.~II of  Ref.~\cite{WiMoJe2022},
one can write the nonrelativistic Hamiltonian as 
follows,
\begin{multline}
H_0 = \omega_\plus \left( a_\plus^\dagger \, a_\plus + \frac12 \right) 
- \frac{e}{2m} \,
(1 + \kappa) \, \vec\sigma\cdot \vec B \\
+ \omega_z \, \left( a_z^\dagger \, a_z + \frac12 \right) 
- \omega_\minus \left( a_\minus^\dagger \, a_\minus + \frac12 \right) \,.
\end{multline}
Here, $a_\plus^\dagger$ and $a_\plus$ are the raising 
and lowering operators for the cyclotron 
motion, $a_z^\dagger$ and $a_z$  pertain to the axial motion,
and $a_\minus^\dagger$ and $a_\minus$ are the raising and lowering operators
for the magnetron motion. 
All of there separately fulfill the algebraic relations 
of a harmonic oscillator~\cite{BrGa1986}.
The bound-state energy is
\begin{align}
\label{E0}
E_{k \ell n s} =& \; \omega_c (1+\kappa) \, \frac{s}{2}
+ \omega_\plus \left( n + \frac12 \right)
\nonumber\\[0.1133ex]
& \; + \omega_z \left( k + \frac12 \right)
- \omega_\minus \left( \qq + \frac12 \right) \,,
\end{align}
where the generalized cyclotron and magnetron frequencies 
$\omega_\plus$ and $\omega_\minus$ are
\begin{align}
\label{defomegaplus}
\omega_\plus =& \; \frac12 \, \left( \omega_c +
\sqrt{ \omega_c^2 - 2 \omega_z^2 } \right) \,,
\\[0.1133ex]
\label{defomegaminus}
\omega_\minus =& \; \frac12 \, \left( \omega_c -
\sqrt{ \omega_c^2 - 2 \omega_z^2 } \right)
\approx \frac{\omega_z^2}{2 \omega_c} \,.
\end{align}
It is of note that the excited states 
can be raised from the ground state
via the raising operators,
\begin{align}
\label{psidef}
\psi_{k \qq n s}(\vec r) = & \;
\frac{\left( a^\dagger_z \right)^k}{\sqrt{k!}} \,
\frac{\left( a^\dagger_\minus \right)^\qq}{\sqrt{\qq !}} \,
\frac{\left( a^\dagger_\plus \right)^n}{\sqrt{n!}} \,
\psi_0(\vec r) \, \chi_{s/2} \,,
\\[0.1133ex]
\chi_{1/2} =& \; \left( \begin{array}{c} 1 \\ 0 \end{array} \right) \,,
\qquad
\chi_{-1/2} = \left( \begin{array}{c} 0 \\ 1 \end{array} \right) \,,
\end{align}
where the $\chi_{s/2}$ denote fundamental spin-up and 
spin-down spinors.
For the ground state (normalized to unity), one obtains
\begin{align}
\psi_0(\vec r) =& \;
\sqrt{ \frac{ m \sqrt{ \omega_c^2 - 2 \omega_z^2 } }{ 2 \pi } } \,
\exp\left( - \frac{m}{4} \sqrt{ \omega_c^2 - 2 \omega_z^2 } \, \rho^2 \right)
\nonumber\\[0.1133ex]
& \; \times \left( \frac{ m \omega_z }{\pi} \right)^{1/4} \,
\exp\left( - \frac12 m \omega_z z^2 \right) \,,
\end{align}
and with Eq.~\eqref{psidef}, one may obtain an
arbitrary excited state symbolically~\cite{Wo1999}.
However, one should note that, while the 
raising operators for the cyclotron, 
axial and magnetron motions mutually commute, they 
still act on the same coordinates $x$, $y$ and $z$
of the ground-state wave function.

It is a further peculiarity of the quantum 
cyclotron spectrum that, while there are
no continuum states and all (bound) 
energy eigenstates can be 
normalized to unity, the nonrelativistic 
spectrum is still not bounded from below, 
in view of the fact that one can lower (ever so slightly) 
the energy by moving away from the symmetry axis 
of the trap, {\em i.e.}, by raising the magnetron
quantum number.
An illustration of a specific state is provided in Fig.~\ref{fig1}.

%
%
\section{Relativistic Corrections}
\label{sec3}

The Foldy--Wouthuysen program, which leads to 
a consistent formulation of the relativistic
corrections, has been outlined in Refs.~\cite{BrGa1986,WiMoJe2022},
for the physical system at hand.
The algebra becomes very lengthy, and computer symbolic 
programs~\cite{Wo1999} constitute a tremendous help
in the formulation of the problem.
In order to expand the operators perturbatively,
one may define generalized fine-structure 
constants [see also Eq.~\eqref{alphac}] which are proportional to the 
square root of the cyclotron and magnetron
frequencies, divided in each case by the 
electron mass (in natural units); one
consults Eqs.~(35) and (36) of Ref.~\cite{WiMoJe2022}. 
In the SI unit system, one would divide the 
the quantum excitation energies of the 
cyclotron and axial motions by the electron rest-mass
energy, and take the square root,
in order to obtain the generalized 
fine-structure constants for the cyclotron and axial
motions. For typical parameters, these generalized
fine-structure constants turn 
out~\cite{WiMoJe2022} to be a lot smaller than the 
quantum electrodynamic fine-structure constant $\alpha_{\rm QED} 
\approx 1/137.036$, a fact which accounts for the comparatively
weaker binding of an electron in a Penning trap,
as compared to, say, a hydrogen atom~\cite{JeAd2022book}.

In analogy to hydrogen bound states~\cite{ItZu1980,JeAd2022book},
the leading relativistic corrections, denoted here as $\delta E$,
to the bound-state energy are obtained in in the second order
of the frequencies $\omega_c$ and $\omega_z$. 
One takes into account the relativistic effects in the 
first order of perturbation theory.
The result is quoted here from Eq.~(90) of
Ref.~\cite{WiMoJe2022} and reads
\begin{multline}
\label{E1}
\delta E = - \frac{ \left[
\frac{ \omega_\plus^2
\left(n+\tfrac12\right) + \omega_\minus^2
\left(\qq + \tfrac12\right) }{\omega_\plus - \omega_\minus}
+ \frac{\omega_z}{2} (k+ \tfrac12)
+ \frac{\omega_c s}{2} \right]^2
}{2m}
\\
- \frac{\omega_z^4
\left[ (n + \tfrac12) (\qq + \tfrac12) + \tfrac14 \right]
}{4m (\omega_\plus - \omega_\minus)^2}
- \frac{ \omega_z^2 }{16 m}
\left[ \left(k + \tfrac12\right)^2 + \tfrac34 \right].
\end{multline}
It is perhaps a surprise to note
that, in the long time between the 
publication (Ref.~\cite{BrGa1986}, in the year 1986)
and the recent investigation reported
in Ref.~\cite{WiMoJe2022} (in the year 2022),
no independent complete investigations 
of the relativistic corrections to the energies
of quantum cyclotron states seem to have been carried out,
to the best of our knowledge.
A comparison to Ref.~\cite{BrGa1986}
verifies a necessary modification of the 
results given in Eq.~(7.48) of Ref.~\cite{BrGa1986},
which can be traced to the replacement
\begin{equation}
\underbrace{ (n + \tfrac12) (\qq + \tfrac12) - \tfrac14 }_%
{\mbox{[Eq.~(7.48) of Ref.~\cite{BrGa1986}]}} \to
\underbrace{(n + \tfrac12) (\qq + \tfrac12) + \tfrac14 }_%
{\mbox{[Result obtained in Ref.~\cite{WiMoJe2022}]}} \,.
\end{equation}
Phenomenologically, the modification
is small, as it pertains to a term which 
is proportional to the fourth power 
of the (comparatively small) {\em axial}
(rather than cyclotron) frequency;
it is still a significant effect which could be
important for trap geometries with 
enhanced numerical values of the axial
frequency (such an arrangement could potentially
reduce systematic effects connected to 
considerable spreading of the electron
wave packet in the axial direction, which is 
observed for low numerical values of the 
axial frequency~\cite{FaGa2021prl,FaGa2021pra}.

%
%
\section{Leading Self--Energy}
\label{sec4}

In order to explore the rich structure
of the operator algebra governing 
quantum cyclotron states, 
let us investigate the leading
spin-independent term of the bound-state self energy.
We first remember the leading-logarithmic 
term in the self-energy 
of bound states in hydrogenlike systems
with charge number $Z$, according to Eqs.~(4.316), 
(4.320) and (4.322) of Ref.~\cite{JeAd2022book}.
In the nonrecoil limit of an infinite nuclear 
mass, the leading-logarithmic energy shift 
$E_L$ is given by
\begin{multline}
\label{LL}
E_L = \frac{2 \alpha}{3 \pi} \left< \frac{\vec p}{m} \,
(H_S - E_S) \, \frac{\vec p}{m} \right> \, \ln[(Z\alpha)^{-2}
\\
= \frac{4 \alpha}{3 \pi} \frac{(Z\alpha)^4 m}{n^3} 
\delta_{l 0} \, \ln[(Z\alpha)^{-2}] \,,
\end{multline}
where $\vec p$ is the momentum operator, 
$H_S$ is the Schr\"{o}dinger Hamiltonian,
$E_S$ is the bound-state energy, $l$ is the 
orbital angular momentum (in this article, $\ell$
is reserved for the magnetron quantum number),
and $n_S$ is the Schr\"{o}dinger principal 
quantum number (here in this article, 
$n$ is reserved for the cyclotron quantum number).

For quantum cyclotron states, analogous 
considerations~\cite{JeMo2023mag2} lead to the 
result that the spin-independent leading-logarithmic 
correction is proportional to the 
matrix element
\begin{equation}
M = \left< \psi_{k \ell n s} |
\vec \pi ( H_0 - E_{k \ell n s} ) \vec \pi 
| \psi_{k \ell n s} \right> \,,
\end{equation}
where $H_0$ is given in Eq.~\eqref{H0}, 
$E_0$ is given in Eq.~\eqref{E0},
and $\vec \pi = \vec p - (e/2) (\vec B \times \vec r)$
is the kinetic momentum in the magnetic trap
field. 

In order to render our task manageable,
we shall work in the approximation
\begin{equation}
\omega_z \approx 0 \,, \qquad
\omega_\plus \approx \omega_c \,, \qquad
\omega_\minus \approx 0 \,,
\end{equation}
in which limit the 
magnetron shift becomes zero, 
and $\ell$ acts as a degeneracy index.
Furthermore, we shall neglect the 
anomalous magnetic moment for simplicity,
and replace $\kappa \to 0$.
Under these approximations, we
can easily establish that
\begin{equation}
\frac{ \vec \pi^{\,2} }{2m} \, \psi_{k \ell n s} \approx
\omega_c \left( n + \frac12 \right) \, \psi_{k \ell n s} \,.
\end{equation}
The unperturbed Hamiltonian becomes
\begin{equation}
H_0 \approx \frac{\vec\pi^2}{2 m} -
\frac{e}{2 m} \vec \sigma \cdot \vec B \,,
\end{equation}
and the unperturbed energy is approximated as 
\begin{equation}
E \approx \omega_c \, \left( n + \frac{s}{2} + \frac12 \right) \,.
\end{equation}
Let us evaluate the matrix element $M$, 
by first showing that $M$ can be written 
in the form
\begin{equation} 
M = X \,,
\qquad
X = \frac{1}{2 m} \left< \vec \pi \left(
\vec\pi^2 \right) \vec \pi \right> -
\frac{1}{2 m} \left< \vec \pi^2 \, \vec\pi^2 \right>  \,.
\end{equation}
The transformations are nonobvious.
One first shows that
\begin{multline} 
M = \left< \vec \pi ( H_0 - E_{k \ell n s} ) \vec \pi \right> 
\\
= \left< \vec \pi \left( 
\frac{\vec\pi^2}{2 m} -\frac{e}{2 m} 
\vec \sigma \cdot \vec B - E_{k \ell n s}
\right) \vec \pi \right> \\
= \left< \vec \pi \,
\frac{\vec\pi^2}{2 m} \, \vec \pi 
+ \left( \frac{\omega_c s}{2}
- E_{k \ell n s} \right) \, \vec \pi^2 \right> 
\\
= X + 
\left< \frac{ \vec \pi^2 \, \vec\pi^2 }{2m} \right>
+ \left< \left( \frac{\omega_c s}{2}
- E_{k \ell n s} \right) \, \vec \pi^2 \right> \,.
\end{multline} 
With the help of the relation
\begin{equation}
\left< \vec \pi^2 \, \vec\pi^2 \right> =
\left[ 2 m \omega_c (n+\tfrac12) \right]^2 \,,
\end{equation}
one now shows that
\begin{multline} 
\left< \frac{ \vec \pi^2 \, \vec\pi^2 }{2m} \right>
+ \left< \left( \frac{\omega_c s}{2}
- E_{k \ell n s} \right) \, \vec \pi^2 \right> 
= \frac{ \left[ 2 m \omega_c (n+\tfrac12) \right]^2 }{2 m}
\\
+ \left[ \omega_c \, \frac{s}{2} 
- \omega_c \, \left( n + \frac{s}{2} + \frac12 \right) \right] 
2 m \omega_c \, \left( n + \frac12 \right) = 0 \,.
\end{multline} 
We are still not done. Namely, $M$ can be written in terms
of Cartesian components, denoted with superscripts,
and with the help of the Einstein summation
convention, as 
\begin{multline}
M = X = \frac{1}{2 m} \left< \vec \pi \left(
\vec\pi^2 \right) \vec \pi \right> -
\frac{1}{2 m} \left< \vec \pi^2 \, \vec\pi^2 \right> 
\\
= \frac{1}{2 m} 
\left( \left< \pi^i \, \pi^j \, \pi^j \, \pi^i \right> -
\left< \pi^i \, \pi^i \, \pi^j \, \pi^j \right> \right)
\\
= \frac{1}{2 m}
\left< \pi^i \, [ \pi^j \, \pi^j, \, \pi^i ] \right>  \,.
\end{multline}
For the commutator,
one trivially has the relation
$[ \vec \pi^{\,2}, \pi^i ] = 
[ \pi^j \, \pi^j, \pi^i ] = 
[ \pi^j , \pi^i ] \, \pi^j +
\pi^j \, [ \pi^j , \pi^i ] $,
and we recall the $i$th Cartesian 
component of the kinetic momentum as
$\pi^i = 
p^i + m \omega_c (\hat{e}_z \times \vec r)^i$.
One easily establishes the commutator relation
\begin{equation} 
[ \pi^i, \pi^j ] = 
\ii e \, \epsilon^{ijk} B^k \, ,
\end{equation} 
where the Levi--Civit\`{a} tensor is $\epsilon^{ijk}$,
and infers that
$[ \vec \pi^{\,2}, \pi^i ] =
2 \ii e \, \epsilon^{ikj} B^k \pi^j$.
So,
\begin{multline}
M = X = 
\frac{1}{2 m}
\left< \pi^i \, [ \vec\pi^{\,2}, \, \pi^i ] \right> 
= \frac{1}{2 m} \left< \pi^i \, 
2 \ii e \, \epsilon^{ikj} B^k \pi^j \right>
\\
= \frac{1}{2 m} \ii e \, \epsilon^{ikj} B^k
\left< [ \pi^i, \, \pi^j ] \right>
= \frac{e^2 }{2 m} \, \epsilon^{ijk} 
\, \epsilon^{ij\ell} \, B^k \, B^\ell 
\\
= \frac{e^2 }{2 m} \, 2 \delta^{k \ell} \, B^k \, B^\ell 
= \frac{e^2 }{m} \, \vec B^2 
= m \, \omega_c^2 \,.
\end{multline}
The step from the first to the second line in the 
above derivation is somewhat nontrivial, 
but it become obvious after a careful consideration.

Here is a decisive difference between the 
Coulombic and quantum-cyclotron bound states:
The matrix element given in Eq.~\eqref{LL}
is nonvanishing only for $S$ states,
where $l = 0$, and it is proportional to 
$1/(n_S)^3$, commensurate with the 
tendency of Coulombic bound states to 
become {\em less relativistic} for higher 
principal quantum numbers $n_S$, 
and thus, undergo less pronounced 
radiative energy shifts.
For the quantum cyclotron problem, if we add 
the anomalous-magnetic-moment term,
which also constitutes a contribution
from the self-energy (which includes the 
vertex corrections induced by the magnetic 
field), the result is, in leading order,
\begin{equation}
\label{alphac}
E_{\rm SE} \sim 
\frac{\alpha}{4 \pi} \, s \, \alpha_c^2 m + E_c \,,
\qquad
\alpha_c = \sqrt{\frac{\omega_c}{m}} \,,
\end{equation}
where $\alpha_c$ is the {\em cyclotron fine-structure
constant}.
The first term (anomalous magnetic moment)
is proportional to 
$s = \pm 1$, where $s$  denotes the spin orientation
[see Eq.~\eqref{s}].
The constant shift $E_c$, on the 
other hand, which is the leading logarithm of the 
self-energy contribution to the Lamb shift,
\begin{equation}
\label{Ec}
E_c = \frac{\alpha}{\pi} \, \alpha_c^4 m \,
\frac23 \ln\left(\alpha_c^{-2} \right) \,,
\end{equation}
is proportional to the matrix element $M$
and is independent of the quantum numbers.
Here, we have show the state-independence
of the matrix element $M$. 
The steps leading from $M$ to 
the leading-logarithmic term $E_c$, 
are explained in further detail in Ref.~\cite{JeMo2023mag2}:
we mention that $E_c$ contributes
to the $\calT_2$ term given in Eq.~(58b)
of Ref.~\cite{JeMo2023mag2}.

%
%
\section{Conclusions}
\label{sec5}

In order to conclude this brief 
investigation, we summarize the 
main findings: In Sec.~\ref{sec2},
we have explored a few peculiarities,
and idiosyncrasies perhaps,
of the algebra of quantum cyclotron states.
The structure of the states has been given in 
Eq.~\eqref{psidef}.
This structure has been further analyzed in
Ref.~\cite{Je2023mag1}, and generalized
for relativistic states in Ref.~\cite{JeMo2023mag2}.
It is crucial to observe that the 
raising operators $a_\plus^\dagger$, $a_\minus^\dagger$,
and $a_z^\dagger$ all operate on the 
same coordinates of the wave function,
but still commute (as operators) 
with each other~\cite{BrGa1986,WiMoJe2022}.

Regarding the relativistic corrections
(Sec.~\ref{sec3}),
we review the most important results,
originally obtained in Refs.~\cite{BrGa1986,WiMoJe2022},
Finally, we obtain a somewhat surprising 
result, on the basis of a completely
analytic calculation, in Sec.~\ref{sec4}:
Namely, quantum-cyclotron states
undergo a radiative shift, 
which, in the leading logarithmic approximation,
is independent of the quantum numbers 
of the quantum cyclotron state
(but not in higher orders, see Ref.~\cite{JeMo2023mag2}).
The leading effect is given in Eq.~\eqref{Ec}.
Indeed, this leading-logarithmic term
is state-independent
(constant) for all quantum cyclotron states.
Electrons bound in Penning traps 
thus experience a shift of their 
physical mass, proportional to 
$B^8 \ln( |e| B/m)$, in view of the 
dressing by the trap's magnetic field.
The result is given in Eq.~\eqref{Ec},
where the physical mass shift is
$\delta m_{\rm phys} = E_c \propto 
\alpha_c^4 \ln(\alpha_c^{-2}) 
\propto B^8 \ln( |e|B/m )$.
This mass shift acts as a dynamically induced 
correction to the electron mass,
induced by the external conditions (the magnetic
field) of the Penning trap. In order to clarify its physical 
interpretation, we recall that the 
mass counter term in the self-energy 
of free electrons can be interpreted as the contribution
of the electromagnetic self-interaction of the 
electron to its mass 
(for a comprehensive
discuss, see Chap.~10 of Ref.~\cite{JeAd2022book}).
It enters the mass term in the electron propagator
via an infinite summation of self-energy
insertions into the Feynman propagator
of the electron [see Eqs.~(10.271)--(10.275) of 
Ref.~\cite{JeAd2022book}]. 
Constant terms in the self-energy of quantum cyclotron 
states enter the electron mass term
in the electron propagator by the same 
summation of perturbation-theoretical 
contributions (Feynman diagrams).
The electron mass (inertial as well as
gravitational) is thus predicted to shift,
albeit minimally, for the quantum cyclotron 
states in a Penning trap,
according to Eq.~\eqref{Ec}.

%
%
\section*{Acknowledgments}

This research was supported 
by NSF grant PHY--2110294.

%
%
\section*{Data Availability Statement}

No Data associated in the manuscript.

\end{document}